\documentclass[fleqn,twoside]{article}
\usepackage{espcrc2}
\usepackage{graphicx}
\usepackage[figuresright]{rotating}

\newcommand{\AmS}{{\protect\the\textfont2
  A\kern-.1667em\lower.5ex\hbox{M}\kern-.125emS}}
\title{Quenched scalar-meson correlator with Domain Wall Fermions}

\author{S. Prelovsek \address[MCSD]{Physics Department, Brookhaven National Laboratory, Upton, NY  11973-5000, USA\vspace*{-0.2cm}}, K. Orginos\address[MCSD]{RIKEN BNL Research Center,  Upton, NY  11973-5000, USA}\\RBC Collaboration\thanks{
Talk presented by S. Prelovsek. \newline
We thank RIKEN, Brookhaven National Laboratory and the U.S.\
Department of Energy for providing the facilities essential for the
completion of this work.}}%

\begin{document}

\begin{abstract}
We study  the $\bar qq$ singlet and non-singlet scalar-meson masses 
using domain wall fermions and the quenched appro\-xi\-ma\-tion. 
The singlet mass is found to be smaller than the 
non-singlet mass and indicates that the lowest singlet 
meson state could be lighter than $1$ GeV.   
The two-point functions for very small quark masses are compared 
with expectations from  the small-volume chiral perturbation theory 
and the presence of fermionic zero modes. 

\vspace*{-0.5cm}  

\end{abstract}

\maketitle

\section{Introduction}

\vspace*{-0.3cm}

The nature of the lightest scalar mesons is not well 
understood. The observed $\sigma$ resonance $0^{++}$ 
with  $m\!\simeq \! 480$ MeV and $\Gamma\!\simeq\! 320$ MeV 
\cite{E791} can be  either a manifestation of the $\pi\pi$ 
scattering or  a distinct $0^{++}$ state. Recent lattice 
simulations explore whether the lightest 
singlet $0^{++}$ state is  a $q\bar q$ 
state, a mixture  of $\bar q q$ and glueball
 or a $\bar q^2 q^2$ state \cite{nakamura,michael,jaffe,weingarten}. 
A dynamical  simulation  of  $\bar q q$ in Ref. \cite{nakamura}  
 indicates  $m_\sigma^{lat}\!\simeq\! m_\rho^{lat}$, 
 while Ref. \cite{michael} finds $m_\sigma^{lat}$ as low as $m_\pi^{lat}$   
indicating possible restoration of chiral symmetry at finite volume. 
In quenched studies,  $\bar q q$-glueball mixing renders the 
lowest scalar mass both  below $1$ GeV \cite{michael} and above $1$ 
GeV \cite{weingarten}. 

The lightest observed non-singlet scalar states are  $a_0(980)$ and 
$a_0(1450)$. Previous lattice simulations indicate that lowest 
scalar $(\bar qq)_{I=1}$  state  is heavier than $1$ GeV 
\cite{michael,weingarten,bardeen}. This opens
 the question on the nature of the $a_0(980)$, proposed to be  
 a $\bar q^2 q^2$ state in \cite{jaffe}. Dynamical simulation  
\cite{milc} observes an indication for the  $a_0\to \eta^\prime \pi$ 
decay.      

 Those  simulations were using Wilson 
\cite{nakamura,michael,jaffe,weingarten} and Kogut-Susskind 
\cite{milc} fermions. 
We re-examine the light scalar spectrum   
using  Domain Wall fermions (DWF), which  allow 
us to study lighter quark masses. Good chiral properties of 
DWF  are especially welcome in the study of the 
$\sigma$ meson, which is intimately related to chiral symmetry breaking. 

A light $\!\sigma\!$ could play an important 
role as an intermediate 
state in $K\!\!\to\!\! \sigma\!\!\to\!(\pi\pi)_{I=0}$, for example.

\vspace*{-0.3cm}

\section{Lattice simulation}

\vspace*{-0.3cm}

We use the quenched approximation together with 
DWF  ($L_s\!=\!16$) and RG-\-impro\-ved gauge action (DBW2).  
This  renders the  chiral symmetry breaking parameter 
$m_{res}$ smaller than 
$1$  MeV. Our lattice parameters are shown in Table 1.

\vspace*{-0.9cm}

\begin{table} [h]
\begin{center}
\begin{tabular}{c c c c c c } 
\hline
$a^{-1}$ & $V$ &  conf. & $m_\pi a$ & $m_0$ [MeV] \\
\hline
 $2$ GeV  & $16^3 32$ & $300$ & $[0.18,~0.26]$ & $800\!\pm\! 40$    \\
$1.3$ GeV & $16^3 32$ & $100$ & $[0.30,~ 0.9]$& $740\!\pm\! 35$\\
$1.3$ GeV & $8^3 24$ & $400$ & $[0.33,~0.9]$& $700\!\pm\! 25$\\
\hline
\end{tabular}
\end{center}
\end{table}

\vspace*{-1.3cm}
\noindent 
Table 1: Simulation parameters, ranges of $m_\pi a$ and results 
for $m_0$.

\vspace*{0.2cm}

The disconnected quark diagram 
is simulated with the Kuramashi technique \cite{kuramashi}.  
This  technique is applied also for the connected diagram, so that a 
meson has a point source on every lattice site.

\vspace*{-0.2cm}

\section{Connected diagram and$\ $quenching$\ $effects}

\vspace*{-0.3cm}

The connected two-point function exhibits conventional exponential 
decay for $m_q\!>\! m_s$ and gives us  the $a_0$
 mass consistent with  \cite{michael,weingarten}.
\begin{figure}[ht]
\begin{center}
\includegraphics[height=4.5cm,width=6.5cm]{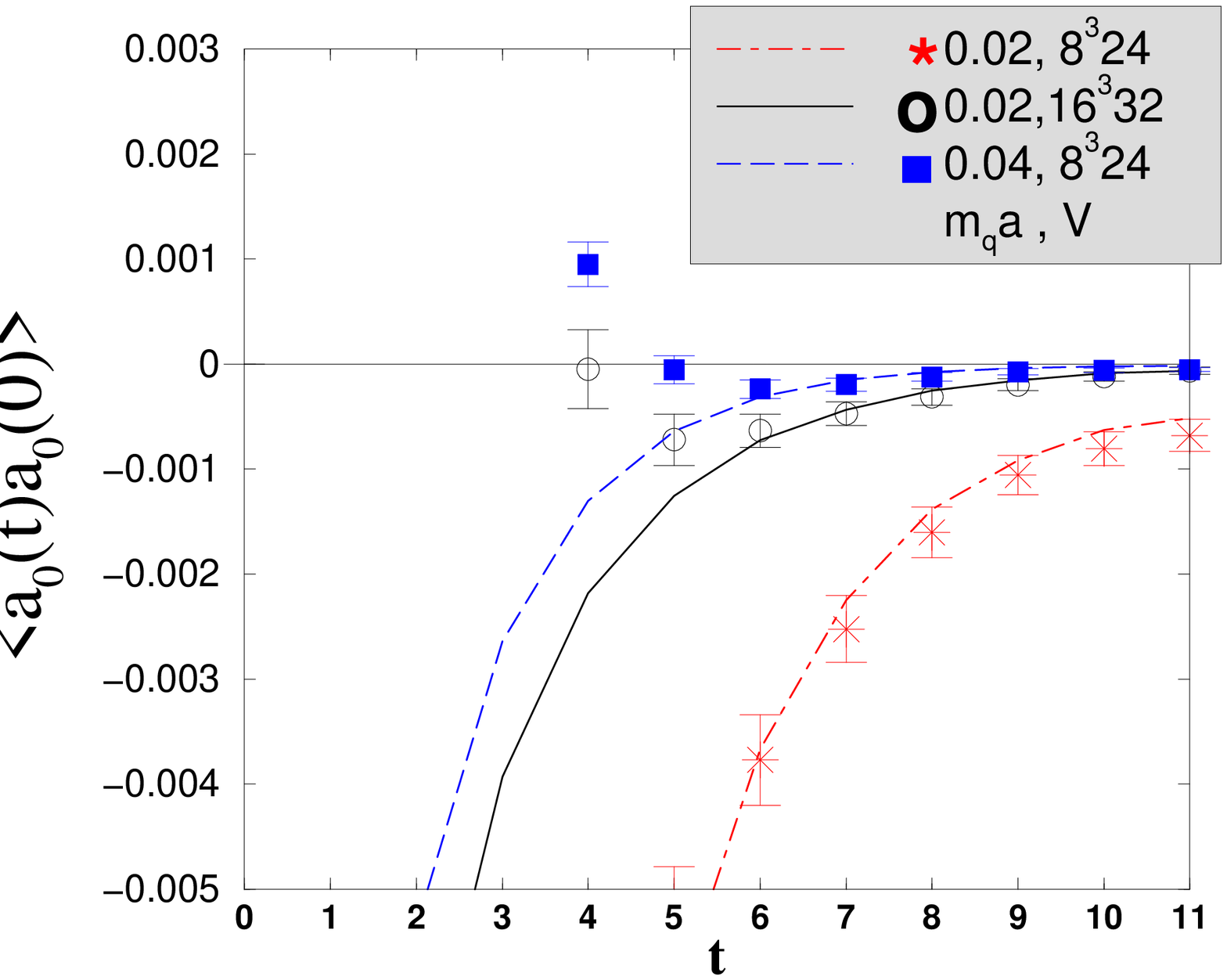}

\vspace*{-1.2cm}

\caption{ Connected scalar two-point function at $a^{-1}\!\!=\!\!1.3$ GeV. 
Lines are $Q\chi PT$ predictions based on the diagram in Fig. 2.}

\vspace*{0.2cm}

\includegraphics[height=1.2cm,width=5cm]{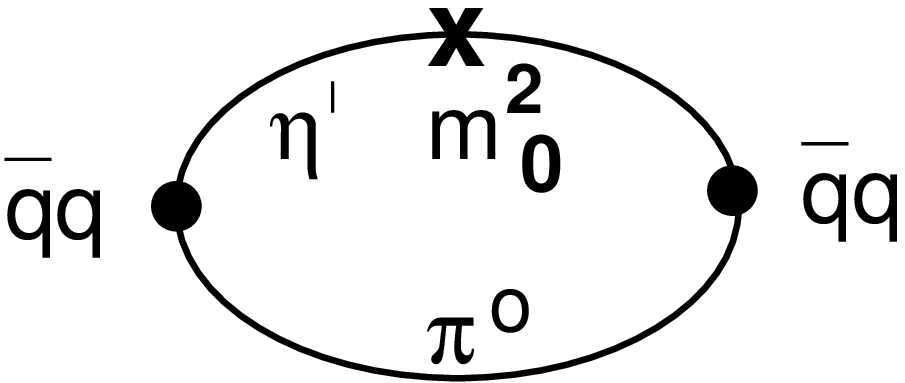}

\vspace*{-1cm}

\caption{This diagram gives the dominant quenched effect 
to connected scalar correlator.}

\vspace*{-1.0cm}

\label{bubble}
\end{center}
 \end{figure}
\begin{figure}[hb!]
\begin{center}

\vspace*{-0.7cm}

\includegraphics[height=3.5cm,width=6cm]{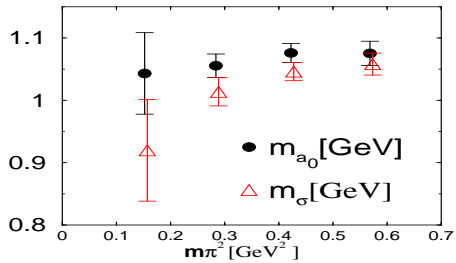}

\vspace*{-1cm}

\caption{Singlet and non-singlet scalar masses for our largest  lattice $16^3\!\times\! 32$ at $a^{-1}\!=\!1.3$ GeV.}

\vspace*{-0.5cm}

\end{center}
\label{msigma}
\end{figure} 
\begin{figure}[htb!!!]
\begin{center}

\includegraphics[height=4.0cm,width=6cm]{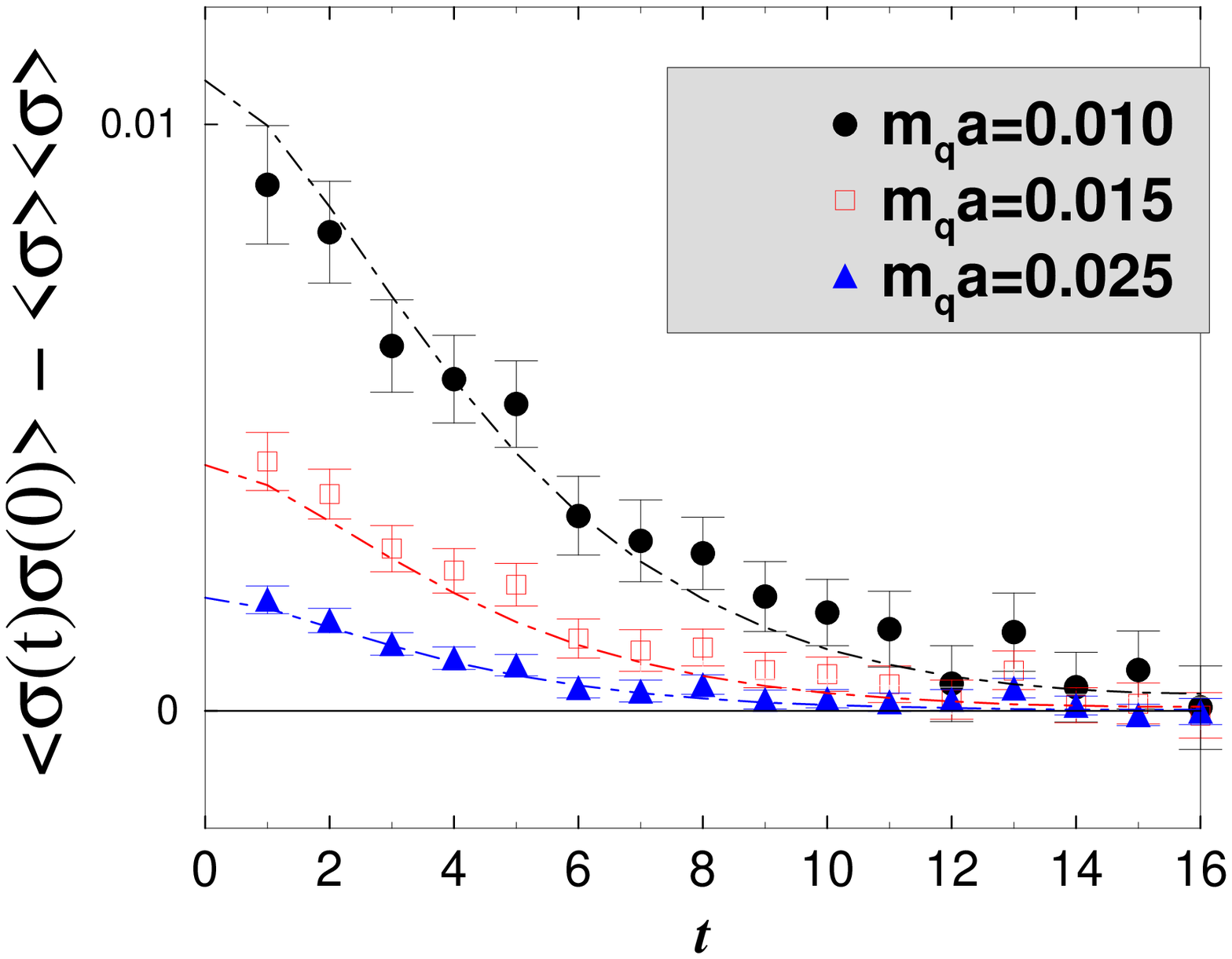}

\vspace*{-1.2cm}

\caption{Disconnected scalar two-point function  at $16^3\!\times\! 32$ and $a^{-1}=2$ GeV together with  the fits to the formula given in Section 4.}

\vspace*{0.3cm}

\includegraphics[height=1.2cm,width=6cm]{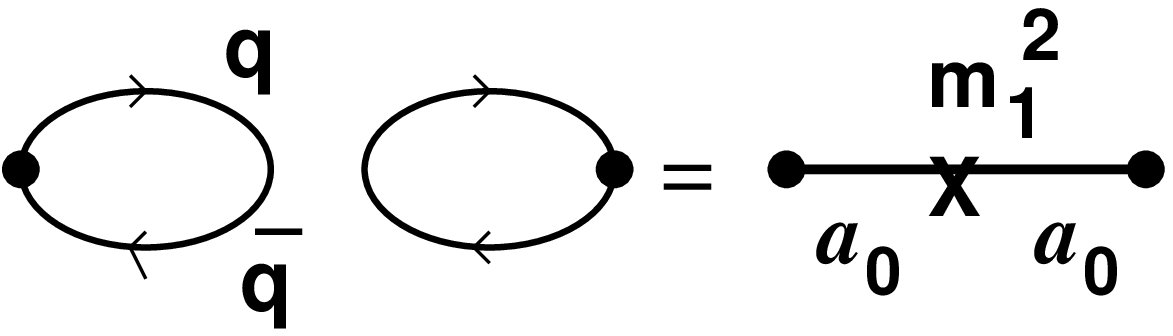}

\vspace*{-1.2cm}

\caption{The disconnected two-point function on the quark and the meson level.}

\vspace*{-1.2cm}

\end{center}
\label{disc}
\end{figure}

 The connected two-point function has a 
striking quenched effect for $m_q<m_s$ \cite{bardeen}
- it is negative at larger times! 
Fig. 1 shows  that the  
negativity is more pronounced for smaller quark masses and smaller volumes.
We  also find that the  correlation function is negative only for non-zero   
topological charge $Q$.  This is 
consistent with the effects of fermionic zero modes, which are not 
suppressed in the quenched approximation  and 
whose contributions are related by \cite{spectrum_RBC}

\vspace*{-0.12cm}

\begin{equation}
\textstyle{
 \!\!\!\!\int \!\! d^4x \langle a_0(x)a_0^\dagger(0)\rangle_{0}\!\!=
\!\!-\int \!\! d^4x \langle \pi(x)\pi^\dagger(0)\rangle_{0}\!\!=-
\frac{|Q|}{Vm_q^2}}.
\label{zeromodes}
\end{equation}

\vspace*{-0.12cm}

The negativity  can 
be understood also within quenched chiral perturbation theory ($Q\chi PT$). 
The small-volume version of $Q\chi PT$ \cite{damgaard} respects the relation (\ref{zeromodes}) and is discussed in the  Section 5. It is applicable when 
 pion wavelength $\lambda_\pi>L$.      

We extract  $m_{a_0}$  using 
 the large-volume version of $Q\chi PT$ where
 $\lambda_\pi\!<\!L$ \cite{bardeen}. 
The dominant negative contribution to $\langle a_0(t)a_0^\dagger(0)\rangle$ 
in this approach arises  
from the $\pi\eta^\prime$ 
intermediate state \cite{bardeen} (Fig. 2). This diagram
 dominates at larger $t$, where the 
 $e^{-m_{a_0}t}$ term is sub-dominant. 
The $q\bar q\!-\!\pi\eta^\prime$ coupling is $m_\pi^2/m_q$. 
The finite volume is taken into account by replacing the 
integral over momenta with a finite sum. 
The value of the $m_0$ insertion is determined from the 
disconnected  pseudoscalar diagram and is 
given in Table 1. 
The value of $\delta\!=\!m_0^2/(24 \pi^2f_\pi^2)\!\simeq\! 0.14\pm 0.04$ 
($f_\pi\! \simeq\!0.13$ GeV) 
obtained from  this $m_0$ is larger than $\delta$ 
obtained from $m_\pi$ \cite{spectrum_RBC} and this discrepancy is currently
under study. We use $m_0^2$ from the hairpin to be consistent with $m_1^2$ 
determined in analogous way below.    
 With this input, the  diagram in Fig. 2 
is  predicted with no free parameters and agrees well with 
our data for different quark masses and volumes, as shown in Fig. 
1.  Surprisingly, the agreement is good also for 
the case of smallest $m_q$   and  $V$, where effects from fermionic 
zero modes are thought to be important.

The complete $Q\chi PT$ description of the 
$\langle a_0 (t)a_0^\dagger(0)\rangle$ correlator incorporates 
pseudoscalar and scalar fields \cite{bardeen}, and  is 
given by the sum of the $a_0$-exchange diagram, 
the  diagram in Fig. 2 and  
their chiral corrections (given in Fig. 8 of \cite{bardeen}). 
The sum of the diagrams  
describes our data very well and 
depends only on two free-parameters: 
the ${a_0}$ mass and its coupling 
$\sqrt{8}f_{a_0} m_\pi^2/m_q$ to the $\bar q q$ 
current\footnote{$m_{a_0}$ and $f_{a_0}$ are denoted by $m_s$ 
and $f_s$ in \cite{bardeen}. }. 
On the lattice with the largest volume, 
which has smaller finite-size and quenching effects (Fig.1), we find  
 $f_{a_0}\!\!\sim \! 0.05$ GeV insensitive to $m_q$, 
while $m_{a_0}$ is shown  in Fig. \ref{msigma}. 
Our result for $m_{a_0}$  is lower than those of 
\cite{michael,weingarten} since we have taken into account the intermediate 
two-pseudoscalar states in the analysis of the two-point 
function
\footnote{The result for $m_{a_0}$ in \cite{bardeen} is larger  
since their value of $f_\pi$ is somewhat higher than the 
physical value, so the chiral corrections are smaller.}. 
Our  result does not exclude the possibility 
of $a_0(980)$ being a $\bar qq$ state.
\begin{figure*}[htb!!!]

\vspace*{-0.2cm}

\includegraphics[height=3.7cm, width=4.7cm]{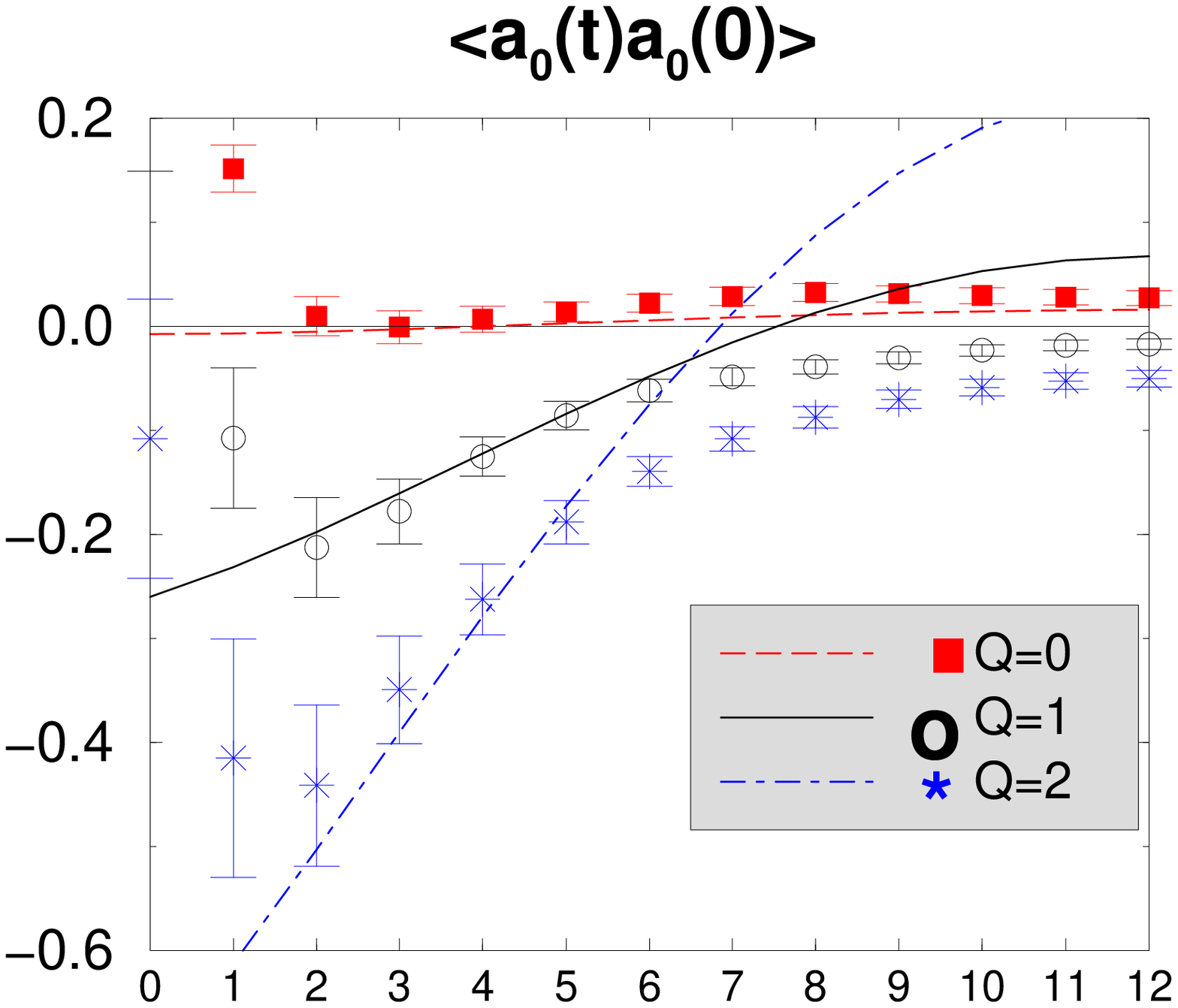}
\hspace*{0.5cm}
\includegraphics[height=3.7cm, width=4.7cm]{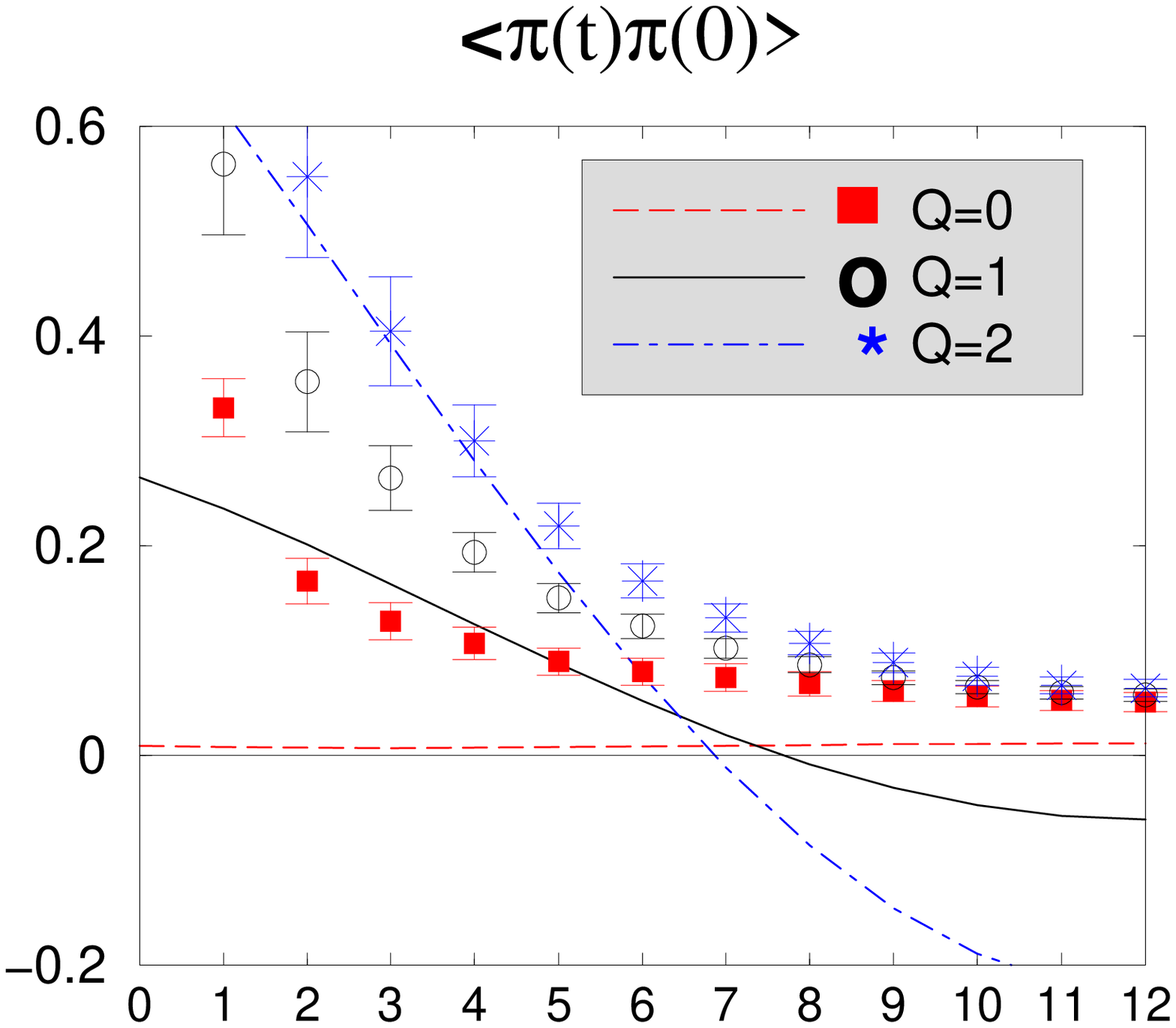}
\hspace*{0.5cm}
\includegraphics[height=3.7cm,width=4.7cm]{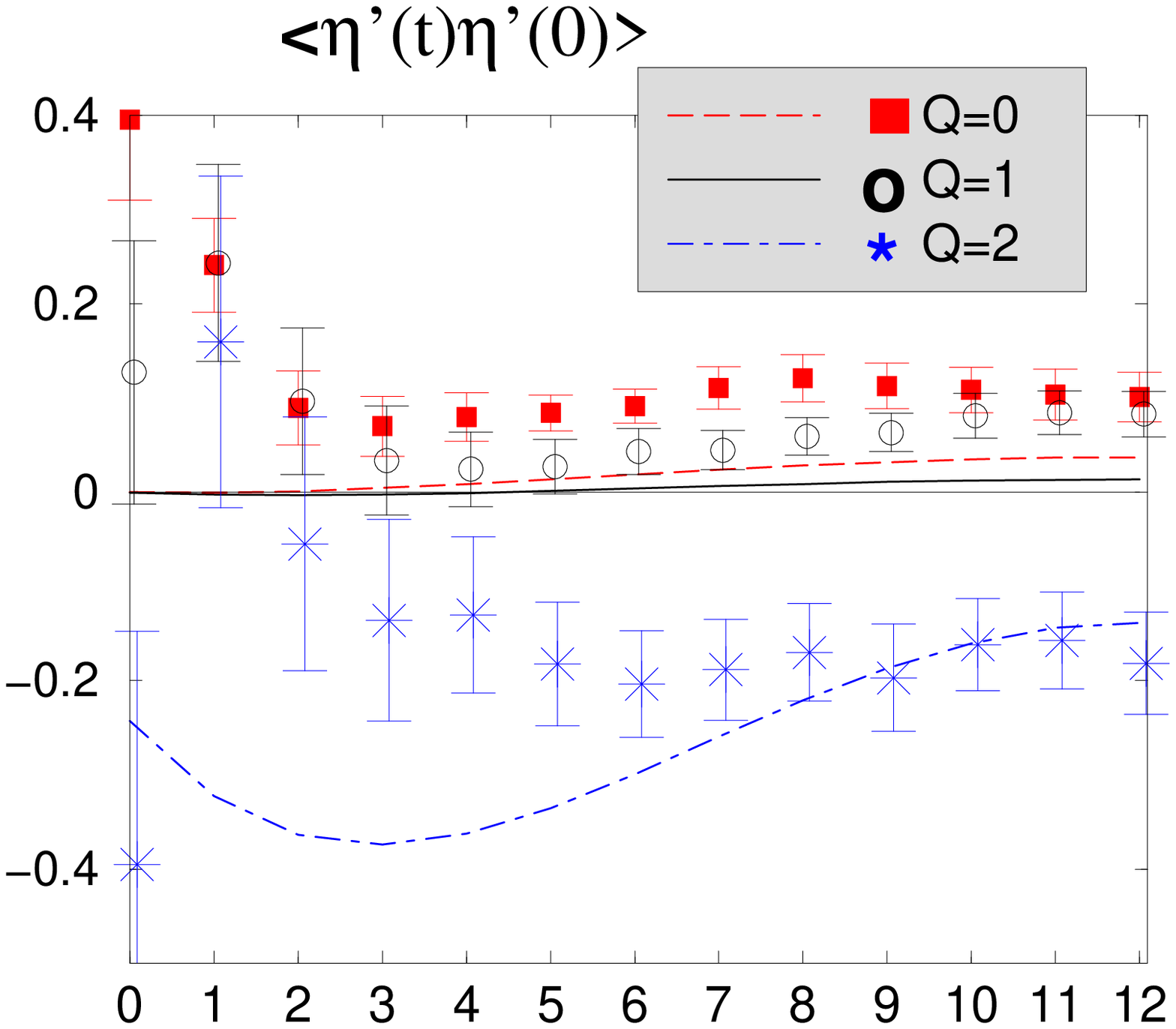}

\vspace*{-1.0cm}

\caption{Small-volume $Q\chi PT$ predictions \cite{damgaard} and our data for $m_q a\!=\!0.005$, $8^3\!\times\! 24$ and $a^{-1}\!=\!1.3$ GeV.  }
\label{damgaard}

\vspace*{-0.4cm}

\end{figure*} 

\vspace*{-0.25cm}

\section{Disconnected diagram and $\sigma$ meson}

\vspace*{-0.3cm}

Our data for the disconnected two-point function  is given in Fig. 4. 
We analyze it in terms of the meson diagram\footnote{
 We approximate $m_1^2$  to be a local coupling since the lightest glueball 
 has a heavy mass of $1.6$ GeV \cite{weingarten}.} in Fig. 5: 
$\langle \sigma^0(t)\sigma^{0}(0)\rangle-\langle \sigma^0\rangle\langle \sigma^{0}\rangle=[\sqrt{8}f_{a_0}m_\pi^2/m_q]^2(-m_1^2)/(p^2+m_{a_0}^2)^2$. The values of $m_{a_0}$ and $f_{a_0}$ are taken  from the connected diagram above. We have  checked also that $m_{a_0}$,  determined from the shape of the disconnected two-point  function, is consistent with $m_{a_0}$ obtained from the connected two-point function.  
   A one-parameter fit to $m_1^2$ gives $m_1^2a^2\!=\!-0.11\pm 0.04$ at $m_\pi a\!=\!0.3003\!\pm\! 0.0019$ for our largest lattice, consistent with results from the other two lattices. The  sum of the diagrams with an arbitrary number of $m_1^2$ insertions in the dynamical theory  gives the singlet scalar mass 
$m_\sigma^2=m_{a_0}^2+m_1^2$. The sign of the disconnected 
diagram in our data has an important physical consequence as it implies that $m_1^2\!<\!0$ and $m_\sigma\!<\!m_{a_0}$. The resulting $m_\sigma$ is shown in Fig. \ref{msigma}.

\vspace*{-0.3cm}

\section{$Q\chi PT$ at small volume}

\vspace*{-0.3cm}

Fig. \ref{damgaard}  shows the comparison between  predictions of small-volume $Q\chi PT$ \cite{damgaard} and our data for two-point functions at fixed topology and very small $m_q$.  The dominance of the 
fermionic zero modes as $V\!\to\! 0$ and $m_q\!\to\! 0$ (Eq. \ref{zeromodes})  is 
 reflected in our data as well as the effective theory. 
Note that $Q\chi PT$ \cite{damgaard} gives a negative prediction for the 
pion correlator at this $V$ for larger $t$,  although a pion correlator with a 
point source and sink should be positive on every configuration by definition.  
This probably indicates that the  expansion parameter $1/(Lf_\pi)$ in  $Q\chi PT$ is not small enough at $8^3\!\times\!  24$ with $a^{-1}\!=\!1.3$ Gee and the comparison has to be performed at larger volume. 

\vspace*{-0.25cm}

\section{Conclusion}

\vspace*{-0.3cm}

We examined the lightest $\bar qq$ singlet and non-singlet scalar meson masses using Domain Wall fermions in the quenched approximation.  
Our conclusions on the scalar spectra 
are based on the lattice with the largest volume and are displayed in Fig. \ref{msigma}. The singlet is lighter than non-singlet - this is an inevitable consequence of the sign of the disconnected two-point function. 
Our results $m_{a_0}=1.04\pm 0.07$ GeV  and $m_{\sigma}=0.9\pm 0.1$ GeV at the 
smallest quark mass 
 indicate that the lightest singlet $\bar qq$ state could be 
lighter than $1$ GeV and   can not rule out that the $a_0(980)$ is a 
$\bar qq$ state.

\end{document}